\documentclass[twocolumn,english,pra]{revtex4-1}
\usepackage[T1]{fontenc}
\usepackage[latin9]{inputenc}
\usepackage{color}
\usepackage{amsmath}
\usepackage{amssymb}
\usepackage{graphicx}
\usepackage{esint}

\makeatletter
\@ifundefined{textcolor}{}
{%
 \definecolor{BLACK}{gray}{0}
 \definecolor{WHITE}{gray}{1}
 \definecolor{RED}{rgb}{1,0,0}
 \definecolor{GREEN}{rgb}{0,1,0}
 \definecolor{BLUE}{rgb}{0,0,1}
 \definecolor{CYAN}{cmyk}{1,0,0,0}
 \definecolor{MAGENTA}{cmyk}{0,1,0,0}
 \definecolor{YELLOW}{cmyk}{0,0,1,0}
 }

\@ifundefined{definecolor}
 {\usepackage{color}}{}
\makeatother

\makeatother

\usepackage{babel}
\begin{document}

\title{Signifying quantum benchmarks for qubit teleportation and secure
communication using Einstein-Podolsky-Rosen steering inequalities}

\author{M. D. Reid}

\affiliation{Centre for Atom Optics and Ultrafast Spectroscopy, Swinburne University
of Technology, Melbourne 3122, Australia}
\begin{abstract}
The demonstration of quantum teleportation of a photonic qubit from
Alice to Bob usually relies on data conditioned on detection at Bob's
location. I show that Bohm's Einstein-Podolsky-Rosen (EPR) paradox
can be used to verify that the quantum benchmark for qubit teleportation
has been reached, without postselection. This is possible for scenarios
insensitive to losses at the generation station, and with efficiencies
of $\eta_{B}>1/3$ for the teleportation process. The benchmark is
obtained, if it is shown that Bob can {}``steer'' Alice's record
of the qubit as stored by Charlie. EPR steering inequalities involving
$m$ measurement settings can also be used to confirm quantum teleportation,
for efficiencies $\eta_{B}>1/m$, if one assumes trusted detectors
for Charlie and Alice.  Using proofs of monogamy, I show that two-setting
EPR steering inequalities can signify secure teleportation of the
qubit state. 
\end{abstract}
\maketitle

\section{Introduction }

Quantum teleportation, the process by which a quantum state is transferred
from one party to another, has inspired countless investigations and
many experiments \cite{tele,telecv theory,teleexp,expcv,danube-1,demart tele,distant matterqubitsproband det,long dist,longdistele,longdistphoton,polzikteleatomightshortdist,teleatom shortdist,enst sw cv exp tele,furhybird,bkcvtele}.
In a real experiment where imperfections will be present, it becomes
necessary to distinguish the process of quantum teleportation from
any other process which can be performed classically. The usual procedure
is to determine the fidelity $F$ of the final teleported state relative
to the initial state. In a classical process, the final sate is created
by a {}``measure and regenerate'' strategy. All such strategies
incur extra noise, so that the fidelity cannot exceed a certain value,
$\mathcal{F}_{c}$ \cite{fidleity,ham}. The figure of merit for quantum
teleportation is a fidelity exceeding $\mathcal{F}_{c}$.

A very important example of teleportation for the purpose of quantum
communication \cite{quantum internet,longdistentcoumm} is the photonic
qubit state teleported over long distances \cite{longdistphoton,danube-1,demart tele,teleexp}.
This quantum teleportation has been realised experimentally using
the original protocol of Bennett et al \cite{tele}. The criticism
has been raised however that these experiments may not give truly
{}``loophole-free'' demonstrations, since the fidelity is calculated
by postselection, i.e. by using only the data observed conditional
on detecting a photon at the teleported location \cite{braunkimblecrit}.
A fundamental issue is that loss will become more problematic where
teleportation distances are large (although the storage of entangled
states using quantum memories may overcome this). It is an interesting
question therefore, to ask what levels of overall efficiency can be
tolerated in order to claim loophole-free quantum teleportation. 

Moreover, the problem of how to demonstrate quantum teleportation\emph{
}is closely linked with how to signify the {}``security'' of the
teleported qubit. If Alice teleports a qubit state to Bob, she may
want to know that the state is teleported uniquely to him, and not
also to another observer, Eve \cite{Hily bclone,hbcopyreg,cerf,clone gm,dag}.
A fidelity $F>2/3$ will signify quantum teleportation, which ensures
that there are not an infinite number of identical copies of the qubit
\cite{clone gm}; the fidelity $F>5/6$ will ensure that any {}``copy''
of Bob's qubit held by Eve will have a degraded fidelity (less than
$5/6$) \cite{Hily bclone,dag}. This knowledge could be used to
evaluate the actual security of a string of qubit values that are
teleported to Bob, by enabling calculation of bounds on Eve's error
rate. Security can be measured in terms of the error rate for any
possible Eve, \emph{or}, more generally, in terms of the maximum number
of Eves that can possess a non-degraded copy of Bob's teleported qubit.
However, for such analyses involving lossy systems, the usual approach
taken to treat {}``no detection'' events leads to an increase in
dimension of the Hilbert space \cite{loopholefreesteering,wittman,bellloophole,Bell,smith},
so that the original fidelity benchmarks which assume qubit systems
are not directly applicable in that case.

In this paper, I present a quite different approach to determining
signatures for quantum teleportation. I show how Bohm's Einstein-Podolsky-Rosen
paradox \cite{epr,bohm epr} can be used to confirm {}``loophole-free''
without postselection the quantum teleportation and quantum security
of a qubit. The result relies on a simple proof of monogamy: Bohm's
EPR paradox cannot be shared among more than a finite number of parties.
Bohm's EPR paradox is an example of the subclass of nonlocality called
{}``\emph{quantum steering}'' \cite{hw-steering-1,hw2-steering-1,Schrodinger},
and the method I propose requires two parties to demonstrate violation
of an {}``EPR steering'' inequality \cite{EPRsteering-1,hw-np-steering-1}.

I focus on the so-called {}``entanglement swapping teleportation
scenario'' \cite{en swp,ent sw dis,enst sw cv exp tele,ent swapexpsteer}.
In that case, Alice's qubit, prior to teleportation to Bob, is entangled
with a qubit of Charlie's. This scenario captures the entire teleportation
process, by including the way the Alice's qubit is locally prepared
from an EPR-type state. Hence, we are able to address the question
raised by Braunstein and Kimble \cite{braunkimblecrit}, as to whether
the zero detection events (if properly accounted for) will detract
from the genuine fidelity of the scheme.

We can establish that the quantum benchmark for teleportation has
been reached, if Bob can demonstrate an EPR paradox, based on his
inferred predictions of Charlie's state. This amounts to Bob {}``steering''
Charlie's system (which may be viewed as a record of the qubit teleported
by Alice) \cite{ent swapexpsteer}. We find that quantum teleportation
is predicted for arbitrary nonzero efficiencies $\eta_{C}>0$ at Charlie's
station, and for an overall teleportation efficiency of $\eta_{B}>1/3$.
These bounds give a sufficient (but not necessary) condition for
loophole-free quantum teleportation. Furthermore, if we assume trusted
detectors at Charlie's station, it would be possible to use the $m$-setting
steering inequalities of Saunders et al \cite{hw-np-steering-1} and
Bennet, Evans et al \cite{loopholefreesteering} to confirm quantum
teleportation at much lower efficiencies, $\eta_{B}>1/m$. 

The level of security of teleported qubit can be determined by the
number of settings $m$ associated with the Bohm EPR steering inequality
that is violated. If Bob demonstrates steering of Charlie's system,
using an $m$-setting steering inequality, then there can be a maximum
of $m-2$ Eve's that can possess identical copies of Bob's qubit.
If Bob demonstrates steering of Charlie's system, using a two-setting
Bohm EPR inequality, then complete monogamy of the violation of the
inequality is guaranteed. I will show that this implies a minimum
noise level for the values of Alice's qubits, as inferred by any independent
third party (Eve). The violation of the two-setting Bohm EPR inequality
is predicted for $\eta_{B}>1/2$, provided the assumption of trusted
detectors is justified at Charlie's location.

I conclude with a brief discussion, pointing out the one-sided device-independent
{[}42,43{]} nature of the protocol that is proposed in this paper.
This means that information is given about the security of the teleported
qubit, regardless of the nature of the devices that could be used
by the parties, Bob and Eve.

\section{Demonstrating Bohm's EPR paradox }

\subsection{A Bohm's EPR paradox criterion\emph{ }}

Let us begin with the question of how to confirm Bohm's EPR paradox.
This is the case where two systems ($A$ and $B$) are prepared in
the Bell-Bohm state \cite{bohm epr,Bell} 
\begin{equation}
|\psi\rangle_{s}=\frac{1}{\sqrt{2}}\{|\uparrow\rangle_{A}|\downarrow\rangle_{B}-|\downarrow\rangle_{A}|\uparrow\rangle_{B}\}\label{eq:bohm}
\end{equation}
The spin outcomes measured at $A$ and $B$ are anti-correlated, if
the same spin component is measured at each system. Bob at $B$ can
make a prediction of \emph{any} Pauli spin component $\sigma_{A}^{\theta}$
of Alice's system $A$, by making a measurement on his spin $\sigma_{B}^{\theta}$.
According to the EPR premises, usually called {}``local realism''
(LR), this implies a predetermination of \emph{each} of Alice's spin
components. In the EPR argument, the predetermined spin components
are represented by an {}``element of reality'', which is a hidden
variable, that defines the spin outcome for Alice's system \emph{precisely},
because Bob's prediction is precise. In the ideal case of Eq. (\ref{eq:bohm}),
the hidden variable values are $1$ or $-1$. There is inconsistency
between the EPR premises and the {}``completeness of quantum mechanics'',
because according to LR, all of Alice's spin components are predetermined
simultaneously, and cannot therefore be given by any quantum mechanical
state.

In practice, Bob cannot infer Alice's spins with perfect accuracy.
We need to know what accuracy will be enough, to deduce an EPR paradox.
One useful approach is to use quantum uncertainty relations \cite{eprr-2}.
For three spins, the variances predicted by quantum mechanics for
any quantum system $ $$A$ are always constrained to satisfy \cite{hoftoth-1}
\begin{equation}
(\Delta\sigma_{A}^{X})^{2}+(\Delta\sigma_{A}^{Y})^{2}+(\Delta\sigma_{A}^{Z})^{2}\geq2.\label{eq:unc}
\end{equation}
On recognising that $ $$\langle(\sigma_{A/B}^{\theta})^{2}\rangle=1$,
we note this quantum uncertainty relation can also be written as the
{}``circle condition'' 
\begin{equation}
\langle\sigma_{A}^{X}\rangle^{2}+\langle\sigma_{A}^{Y}\rangle^{2}+\langle\sigma_{A}^{Z}\rangle^{2}\leq1\label{eq:un2}
\end{equation}
used in Refs \cite{wittman,pra}. By extending the above argument
that relates to perfect correlation and ideal states \cite{rrmp,eprr-2},
we can derive an inequality for a practical test of the Bohm EPR paradox.
We demonstrate Bohm's EPR paradox, if 
\begin{eqnarray}
S{}_{A|B}^{(3)} & = & (\Delta_{inf}\sigma_{A|B}^{X})^{2}+(\Delta_{inf}\sigma_{A|B}^{Y})^{2}+(\Delta_{inf}\sigma_{A|B}^{Z})^{2}\nonumber \\
 &  & \,\,\,\,\,\,\,\,\,\,\,\,\,\,\,\,\,\,\,\,\,\,\,\,\,\,\,\,\,\,\,\,\,\,\,\,\,\,\,\,<2\label{eq:ineqbohmepr}
\end{eqnarray}
Here $\Delta_{inf}\sigma_{A|B}^{X}$ is the {}``inference'' uncertainty
for Bob's prediction of Alice's spin $\sigma_{A}^{X}$. Where there
is a need to specify the second party (in this case $B$) that is
making the inference, we will use the explicit notation $(\Delta_{inf}\sigma_{A|B}^{X})^{2}$,
but otherwise we will write $\Delta_{inf}\sigma_{A|B}^{X}$ as $\Delta_{inf}\sigma_{A}^{X}$.
The {}``inference variance'' is the average conditional variance
\begin{equation}
(\Delta_{inf}\sigma_{A|B}^{X})^{2}=\sum_{\sigma_{B}^{\varphi}=-1,+1}P(\sigma_{B}^{\varphi})\{\Delta(\sigma_{A}^{X}|\sigma_{B}^{\varphi})\}^{2}\label{eq:inf}
\end{equation}
This variance gives the uncertainty of the {}``element of reality''
for $\sigma_{A}^{X}$. Here, $\{\Delta(\sigma_{A}^{X}|\sigma_{B}^{\varphi})\}^{2}$
denotes the variance of the conditional distribution $P(\sigma_{A}^{X}|\sigma_{B}^{\varphi})$.
The inference variances for the spins $\sigma^{Y}$ and $\sigma^{Z}$
are defined similarly. We have written (\ref{eq:inf}) as though the
best possible prediction for Alice's spin $\sigma_{A}^{X}$ will be
given by Bob measuring $ $$\sigma_{B}^{\varphi}$, and that this
choice will give the smallest $\Delta_{inf}\sigma_{A}^{X}$. The specification
of which measurement of Bob's is optimal is irrelevant however for
the criterion. (For simplicity of notation, we assume it is understood
from the context whether we are referring to the spin operator measurements
$\hat{\sigma}_{A/B}^{\theta}$ or the outcomes of those measurements,
and omit the {}``hats'' in the first case.)

When the criterion (\ref{eq:ineqbohmepr}) is achieved, the inferred
uncertainties {}``violate'' the uncertainty principle (\ref{eq:unc}),
if they represent simultaneous descriptions of spin components. For
this reason, the inequality (\ref{eq:ineqbohmepr}) will demonstrate
the incompleteness of quantum mechanics, based on the assumption of
LR \cite{rrmp}. The inequality is thus a sufficient condition for
Bohm's EPR paradox.

The Bohm's EPR paradox inequality is closely related to the steering
inequality used by Wittmann et al to demonstrate loophole-free steering.
The close relationship between the EPR paradox and quantum steering
was pointed out in Refs. \cite{hw-steering-1,hw2-steering-1,EPRsteering-1}.
Since $(\Delta_{inf}\sigma_{A|B}^{\theta})^{2}=1-\langle\sigma_{A}^{\theta}|\sigma_{B}^{\varphi}\rangle^{2}$
where $ $$\langle\sigma_{A}^{\theta}|\sigma_{B}^{\varphi}\rangle$
is the mean of $P(\sigma_{A}^{X}|\sigma_{B}^{\varphi})$, substitution
into (\ref{eq:ineqbohmepr}) yields the equivalent inequality
\begin{equation}
S=T_{X}+T_{Y}+T_{Z}>1\label{eq:steerwi}
\end{equation}
where $T_{X}=\sum_{\sigma_{B}^{\varphi}}P(\sigma_{B}^{\varphi})\langle\sigma_{A}^{X}|\sigma_{B}^{\varphi}\rangle^{2}$
(and similarly for $T_{Y}$ and $T_{Z}$). This is precisely the steering
inequality used by Wittman et al \cite{wittman}.

\subsection{Bohm's EPR paradox without fair sampling assumptions\emph{ }}

The inequality (\ref{eq:ineqbohmepr}) does not take into account
detection losses, where one or both of the particles is not detected.
The usual procedure is to introduce a fair sampling assumption, where
all {}``no detection'' events are ignored. In some recent experiments
that detect (without fair-sampling loopholes) the sort of nonlocality
called {}``quantum steering'', the assumption is made that the detectors
for Alice's particle can be {}``trusted'' \cite{loopholefreesteering,wittman,smith}.
This means that the fair sampling assumption is made asymmetrically,
for Alice's system but not for Bob's.

The Bohm EPR condition (\ref{eq:ineqbohmepr}) can be modified, so
that it will apply without fair sampling assumptions, for either party.
This provides a way to demonstrate {}``loophole-free'' the Bohm
EPR paradox. The original condition (\ref{eq:ineqbohmepr}) is derived
from the quantum uncertainty relation (\ref{eq:unc}) which is valid
only when the outcomes for the measurements are dichotomic ($\pm1$).
This uncertainty relation can be modified, to allow for {}``no detection''
events, that are labelled by an outcome of $0$. This approach of
expanding the Hilbert space has been commonly used to treat the effect
of loss on nonlocality \cite{Bell,bellloophole,wittman,loopholefreesteering,smith}. 

It is convenient to introduce the Schwinger formalism for spins\textcolor{black}{.
This enables a direct analogy with the photonic realisation of spin
measurements, whereby the Stern-Gerlach apparatus is replaced by polarising
beam splitters. Two orthogonal polarisation field modes are defined
at each of two sites $A$ and $B$, and are identified by boson operators
$a_{\pm}$ and $b_{\pm}$, respectively. In the ideal case, the spin
states at $A$ are $|\uparrow\rangle_{A}=|1\rangle_{a+}|0\rangle_{a-}$
and $|\downarrow\rangle_{A}=|0\rangle_{a+}|1\rangle_{a-}$, which
describe a photon in one of the polarisation modes. More generally,
the measurable Schwinger spin observables at $A$ are 
\begin{eqnarray}
S_{A}^{Z} & = & a_{+}^{\dagger}a_{+}-a_{-}^{\dagger}a_{-}\nonumber \\
S_{A}^{X} & = & a_{+}^{\dagger}a_{-}+a_{+}a_{-}^{\dagger}\nonumber \\
S_{A}^{Y} & = & (a_{+}^{\dagger}a_{-}-a_{+}a_{-}^{\dagger})/i\nonumber \\
S_{A}^{2} & = & n_{A}(n_{A}+2)\nonumber \\
n_{A} & = & a_{+}^{\dagger}a_{+}+a_{-}^{\dagger}a_{-,}\label{eq:schwinger}
\end{eqnarray}
where $S_{A}^{2}=(S_{A}^{X})^{2}+(S_{A}^{Y})^{2}+(S_{A}^{Z})^{2}$,
and $n_{A}$ is the total number operator. Similar operators and states
are defined for the system $B$. }

Having established the formalism, we now introduce an uncertainty
relation 
\begin{eqnarray}
(\Delta S_{A}^{X})^{2}+(\Delta S_{A}^{Y})^{2}+(\Delta S_{A}^{Z})^{2} & \geq & \langle n_{A}^{2}\rangle-\langle n_{A}\rangle^{2}+2\langle n_{A}\rangle\nonumber \\
\label{eq:spinun-1}
\end{eqnarray}
which will hold for any quantum state, and which follows from $\langle S^{2}\rangle=\langle n(n+2)\rangle$
and that $\langle S_{X}\rangle^{2}+\langle S_{Y}\rangle^{2}+\langle S_{Z}\rangle^{2}\leq\langle n\rangle^{2}$
\cite{tothsch,pra}. This uncertainty relation can be used to derive
an inequality for the Bohm EPR paradox in non-ideal scenarios.

Specifically, by applying the EPR argument with the quantum uncertainty
relation (\ref{eq:spinun-1}), we see that we will verify an EPR paradox
if 
\begin{eqnarray}
(\Delta_{inf}S_{A}^{X})^{2}+(\Delta_{inf}S_{A}^{Y})^{2}+(\Delta_{inf}S_{A}^{Z})^{2}\nonumber \\
<\langle n_{A}^{2}\rangle-\langle n_{A}\rangle^{2}+2\langle n_{A}\rangle\label{eq:bohmeff}
\end{eqnarray}
Here, we define 
\begin{equation}
(\Delta_{inf}S_{A|B}^{X})^{2}=\sum_{s_{B}^{\varphi}=-1,0,+1}P(s_{B}^{\varphi})\{\Delta(S_{A}^{X}|S_{B}^{\varphi})\}^{2}\label{eq:inf-1}
\end{equation}
in accordance with definition (\ref{eq:inf}). This inequality is
the generalisation of the EPR Bohm paradox condition (\ref{eq:ineqbohmepr})
that accounts for detection inefficiencies, at both sites, and is
the main result of this paper. 

The inequality is also a {}``steering'' inequality, and can be derived
directly from the Local Hidden State formalism established in Refs.
\cite{hw-steering-1,hw2-steering-1}. This proof is presented in the
Appendix. We will use the term {}``EPR steering'' to refer to such
inequalities, that test both the EPR paradox and the nonlocality of
quantum steering \cite{EPRsteering-1}.

\textcolor{black}{With only one photon incident at each site, and
the possibility of {}``no detection'', the possible outcomes for
a given {}``spin'' $S_{A/B}^{Z/Y/X}$ are $+1$, $-1$ and $0$.
Denoting the probabilities for each of these outcomes at site $j$
($j\equiv A,B$) by $P_{+j},$$P_{-j}$ and $P_{0j}$ respectively,
we note that 
\[
\langle n_{j}\rangle=P_{+j}+P_{-j}=\eta_{j}
\]
}where $\eta_{j}$ is the efficiency at the site $j$. We use the
notation $\eta_{A}$ for the efficiency at site $A$, and $\eta_{B}$
for the efficiency at site $B$. 

We can also modify the steering inequality used by Witttman et al
\cite{wittman}, so that it accounts for the inefficiencies at the
{}``trusted site'' of Alice. Since the outcomes for each $S_{A/B}^{\theta}$
are $\pm1$ or $0$ (here $\theta\equiv X,Y,Z$), it is easy to verify
that $\langle(S_{A}^{\theta})^{2}\rangle=\eta_{A}$, and hence that
$(\Delta_{inf}S_{A|B}^{X})^{2}=\sum_{s_{B}^{X}}P(s_{X}^{B})\{\eta_{A}-\langle S_{A}^{X}|S_{B}^{X}\rangle^{2}\}=\eta_{A}-T_{X}$,
where here $\langle S_{A}^{X}|S_{B}^{X}\rangle$ denotes the mean
value of $S_{A}^{X}$ given the result $S_{B}^{X}$. Thus, the Bohm
EPR condition (\ref{eq:bohmeff}) can be written 
\begin{equation}
S=T_{X}+T_{Y}+T_{Z}>\eta_{A}^{2}\label{eq:steerwitloss}
\end{equation}
which is the extension of the steering inequality (\ref{eq:steerwi})
used by Wittman et al \cite{wittman}. If this inequality is satisfied,
then one can confirm a steering of system $A$ by measurements performed
by Bob (at system $B$) \emph{without} the assumption of trusted detectors
at Alice's location.

\subsection{Quantum prediction}

We now ask, for what quantum states and with what degree of loss can
the Bohm EPR paradox criterion be satisfied? Let us assume the system
is in a Werner mixed state:\textbf{ }
\begin{equation}
\hat{\rho}=(1-p_{s})\frac{1}{4}I+p_{s}|\psi\rangle_{S}\,_{S}\langle\psi|\,,
\end{equation}
where $p_{s}$ gives the relative contribution of the Bell state $|\psi_{S}\rangle=\frac{1}{\sqrt{2}}\{|\uparrow\rangle_{A}|\downarrow\rangle_{B}-|\downarrow\rangle_{A}|\uparrow\rangle_{B}\}$
and $I$ is a rotationally symmetric, uncorrelated state proportional
to the identity matrix at each site. The calculations are given in
the Supplemental Material \cite{supp}. We find that for a system
in the Werner state and with detection efficiencies $\eta_{A}$ and
$\eta_{B}$ at each site, the quantum prediction is 
\begin{eqnarray}
 &  & (\Delta_{inf}S_{A|B}^{\theta})^{2}=\eta_{A}\{1-\eta_{A}\eta_{B}p_{s}^{2}\}\label{eq:inprd-1-1}
\end{eqnarray}
where $\theta\equiv X,Y,Z$. The Bohm EPR condition (\ref{eq:bohmeff})
is satisfied when  
\begin{equation}
\eta_{B}>1/(3p_{s}^{2})\label{eq:loss}
\end{equation}
i.e. for $\eta_{B}>1/3$ where $p_{s}=1$ (provided $\eta_{A}>0$).
This efficiency for $\eta_{B}$ (Bob's detection) has been achieved,
in experiments of Wittman et al \cite{wittman}. We note that the
criterion is satisfied \emph{independently} of the value of the efficiency
for Alice's detection, so long as it is nonzero.

\textcolor{black}{The EPR steering inequality (\ref{eq:bohmeff})
is very useful for loophole-free tests, since it applies regardless
of the photon numbers }\textcolor{black}{\emph{actually incident}}\textcolor{black}{{}
on the detectors. This means it can fully account for all spurious
events. This is important where the photon pairs are generated via
parametric down conversion, since then there is always a possibility
of two photon pairs being generated. As these events usually occur
with a very small probability, however, the quantum prediction given
here is valid for most scenarios.}

\section{Signature for qubit quantum Teleportation without fair sampling}

We now address the question of how to apply the Bohm's EPR criterion
to demonstrate quantum teleportation. We begin with a simple proof
of monogamy. If a party $B$ can demonstrate a steering of the party
$A$, by satisfying the Bohm's EPR paradox inequality (\ref{eq:ineqbohmepr})
or its generalisation (\ref{eq:bohmeff}), then there cannot be an
infinite number of other parties that can also do this.

\subsection{Monogamy relations for the EPR steering inequalities }

We define a {}``steering parameter'' that is based on the Bohm EPR
criterion (\ref{eq:bohmeff}): 
\begin{equation}
S_{A|B}^{(3)}=\{(\Delta_{inf}S_{A}^{X})^{2}+(\Delta_{inf}S_{A}^{Y})^{2}+(\Delta_{inf}S_{A}^{Z})^{2}\}/J\label{eq:steerp}
\end{equation}
where $J=\langle n_{A}^{2}\rangle-\langle n_{A}\rangle^{2}+2\langle n_{A}\rangle$.
Then we see that according to (\ref{eq:bohmeff}) EPR steering of
system $A$ by $B$ is obtained when $S_{A|B}^{(3)}<1$. We note
that this inequality involves three observables, and is hence a {}``three-setting
inequality''.

We now prove that a monogamy steering relation holds for the steering
parameter. For any four quantum systems $A$-$D$, it is always true
that 
\begin{equation}
S_{A|B}^{(3)}+S_{A|C}^{(3)}+S_{A|D}^{(3)}\geq3\label{eq:steermon3}
\end{equation}
This result, and a collection of other monogamy results for the EPR
paradox and quantum steering, have been presented and proved in previous
papers \cite{monog,mon2}. The proof is briefly summarised here, for
the sake of completeness.

\textbf{Proof:} The observer at $B$ (Bob) can make the measurement
that gives him the value of Alice's observable $S_{A}^{X}$ with uncertainty
$\Delta_{inf}S_{A|B}^{X}$. The observer at $C$ (Charlie) can make
the measurement that gives the result for Alice's $S_{A}^{Y}$ with
uncertainty $\Delta_{inf}S_{A|C}^{Y}$, \emph{and} the observer at
$D$ can make the measurement that gives the result for Alice's $S_{A}^{Z}$
with uncertainty $\Delta_{inf}S_{A|D}^{Z}$. Since the three observers
can measure simultaneously, the uncertainty relation (\ref{eq:spinun-1})
constrains the variances to be $(\Delta_{inf}S_{A|B}^{X})^{2}+(\Delta_{inf}S_{A|C}^{Y})^{2}+(\Delta_{inf}S_{A|D}^{Z})^{2}\geq J$.
Similarly, $(\Delta_{inf}S_{A|D}^{X})^{2}+(\Delta_{inf}S_{A|B}^{Y})^{2}+(\Delta_{inf}S_{A|C}^{Z})^{2}\geq J$
and also $(\Delta_{inf}S_{A|C}^{X})^{2}+(\Delta_{inf}S_{A|D}^{Y})^{2}+(\Delta_{inf}S_{A|B}^{Z})^{2}\geq J$.
We then see that the monogamy relation (\ref{eq:steermon3}) follows,
upon adding the three inequalities. $\square$ 

The monogamy result (\ref{eq:steermon3}) tells us that, within the
constraints of quantum theory, it is impossible for more than two
parties to (independently) demonstrate the steering of system $A$,
by the procedure of violating the $3$-setting steering Bohm EPR paradox
inequality (\ref{eq:bohmeff}).

\subsection{Quantum teleportation of a qubit}

There is a close relationship between monogamy and quantum no-cloning.
We now turn to the situation where Alice teleports a quantum state
to Bob, via an entanglement swapping protocol. The monogamy of the
three-observable steering inequality (\ref{eq:bohmeff}), as given
by (\ref{eq:steermon3}), will restrict the number of equivalent copies
of Alice's state that can be teleported to different parties. Such
a restriction cannot be achieved by any classical {}``measure and
regenerate'' strategy, since such a strategy would allow an infinite
number of equivalent copies to be regenerated.

\begin{figure}[t]
\begin{centering}
\includegraphics[width=1\columnwidth]{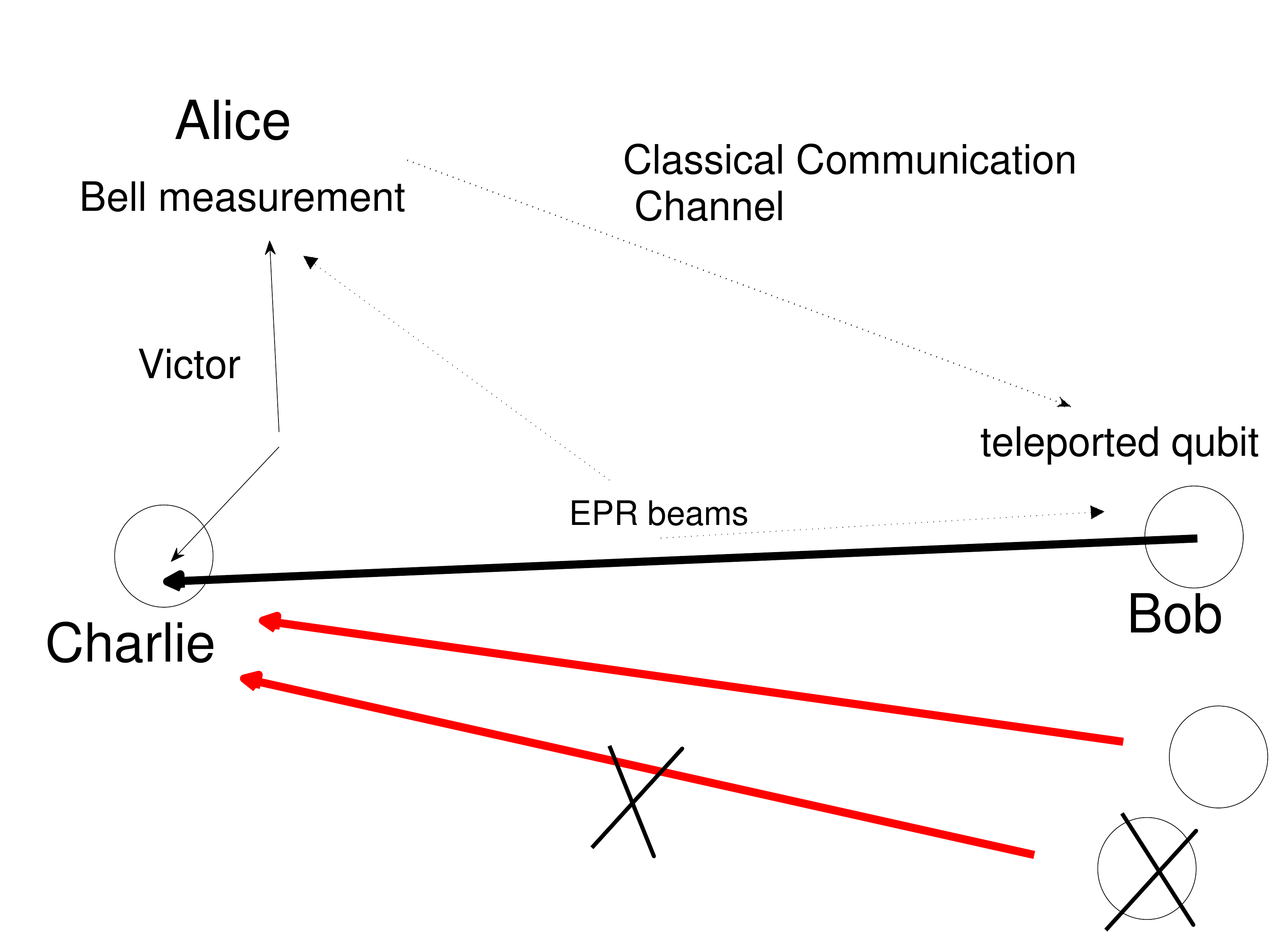}
\par\end{centering}

\caption{\emph{Schematic of verification of quantum teleportation using the
monogamy inequalities.} If Bob can verify steering of Charlie's qubit
system, using a two-observable steering inequality, then no other
party (Eve) can also do this. This excludes the possibility of a clone
of the teleported state, and gives confirmation of secure quantum
teleportation. If Bob can steer Charlie's qubit system using an $m$-setting
steering inequality, then there can be no more than $m-2$ clones
(in the diagram, $m=3$, and the red line indicates {}``Eve''):
the possibility of a classical {}``measure and regenerate'' strategy
is negated, and quantum teleportation confirmed.}
\end{figure}

Let us consider the set-up of the Figure 1, where a Bohm EPR two qubit
state is prepared at the site of Alice, Charlie and Victor. One qubit
is with Charlie. The second EPR qubit is with Victor, and is then
teleported to Bob, by Alice's sending station. After teleportation,
the entanglement is {}``swapped'', and Bob and Charlie share an
entangled EPR state \cite{en swp,ent sw dis,enst sw cv exp tele,ent swapexpsteer}.

In the standard protocol, two EPR beams are prepared in the Bell state
$|\psi\rangle=\frac{1}{\sqrt{2}}\{|\uparrow\rangle_{A}|\downarrow\rangle_{B}-|\downarrow\rangle_{A}|\uparrow\rangle_{B}\}$.
One of these beams is sent to Alice, the other to Bob. Victor and
Charlie prepare an entangled EPR qubit
\begin{equation}
|\phi\rangle=\frac{1}{\sqrt{2}}\{|\uparrow\rangle_{V}^{\theta}|\downarrow\rangle_{C}^{\theta}-|\downarrow\rangle_{V}^{\theta}|\uparrow\rangle_{C}^{\theta}\}\label{eq:epr1}
\end{equation}
and Victor will teleport his qubit to Bob. Victor's qubit is inputed
to a teleportation device, while the correlated qubit remains with
Charlie. Alice performs a Bell measurement on the direct product state
which can be written as a linear combination of the four Bell states
\cite{tele}. If Alice measures the system to be in a Bell state
$|\Psi_{-}\rangle=\frac{1}{\sqrt{2}}\{|\uparrow\rangle_{V}^{\theta}|\downarrow\rangle_{A}^{\theta}-|\downarrow\rangle_{V}^{\theta}|\uparrow\rangle_{A}^{\theta}\}$
$|\Psi_{-}\rangle$, then her result is sent classically to Bob, who
recovers Victor's qubit by performing the identity operation on his
state. The procedure swaps the entanglement between Charlie and Victor
to one between Charlie and Bob, and the final state is the entangled
qubit 
\begin{equation}
\frac{1}{\sqrt{2}}\{-|\uparrow\rangle_{B}|\downarrow\rangle_{C}+|\downarrow\rangle_{B}|\uparrow\rangle_{C}\}\label{eq:epr1-1}
\end{equation}
This description summarises the entire teleportation process, even
when viewed as teleportation of a single qubit, since in practice
the correlated Bell-Bohm EPR state of Victor and Charlie is used to
herald the qubit that is inputed to Alice's sending station \cite{teleexp}. 

If Charlie has detected his qubit (before teleportation so both he
and Alice know the qubit value and angle $\theta$) then we may view
the process as teleportation of a qubit. Victor's input to the sending
station is the anti-correlated eigenstate of definite spin along the
direction $\theta$. If Alice performs a Bell measurement with result
indicating $|\Psi_{-}\rangle$ and transmits this result to Bob, then
he will know that his state is the same state as Victor's input. Alternatively,
if Charlie does not make his measurement, then Bob's state is entangled
with Charlie's state. In that case, if Charlie measures his spin along
the direction $\theta$, then he and Bob will know that Bob's state
is the eigenstate with the anti-correlated spin along the same direction.
Delayed-choice entanglement swapping has been studied in recent experiments
\cite{Peres,ent swapexpsteer}.

\subsection{Signature of quantum teleportation}

In order to claim quantum teleportation, there must be a signature
to indicate that Bob's final teleported state cannot be generated
using any classical strategy. Usually this is done by demonstrating
that the fidelity with Alice's input state is higher than can achieved,
based on any classical {}``measure and regenerate'' protocol. It
has been proved that according to quantum mechanics a high fidelity,
$\mathcal{F}>2/3$, can be only be achieved for a finite number of
copies of a qubit state. This feature is a consequence of the quantum
no-cloning theorem \cite{Hily bclone,fidleity,no clone-1,dag}. Since
for any classical teleportation strategy, an infinite number of identical
copies are possible, the criterion $\mathcal{F}>2/3$ will demonstrate
quantum teleportation for qubit systems. In fact, this fidelity criterion
is a necessary and sufficient condition, and has been used to determine
the optimal teleportation with mixed state qubits \cite{mixed state tele-1}.

Here, I consider a different approach, based on an EPR steering inequality.
Bob measures the value of the teleported qubit along a direction
$\theta$. Charlie communicates classically the value of $\theta$
that he or Victor used to define the qubit, so that Bob knows what
measurement to make. Alice and Charlie know when she sent the information
about the qubit. The experiment involves the quantum state conditional
on her making the Bell measurement and sending the classical signal
to Bob. 

For a sequence of states each with the value of $\theta,$ Bob reports
to Charlie his values, Charlie performs the same measurement of his
spin, and the conditional variance $(\Delta_{inf}\sigma_{C|B}^{\theta})^{2}$
measured. If this is done for three orthogonal selections of $\theta$,
the steering parameter $S_{C|B}^{(3)}$ can be determined. The observation
of steering is given when $S_{C|B}^{(3)}<1$. Suppose Bob and Charlie
verify that Bob has satisfied the EPR steering criterion 
\begin{equation}
S_{C|B}^{(3)}<1\label{eq:telecond}
\end{equation}
Use of the monogamy relation $S_{C|B}^{(3)}+S_{C|D}^{(3)}+S_{C|E}^{(3)}\geq3$
(Eq. (\ref{eq:steermon3})) tells us that there can be no more than
one other party that can also show a steering of Charlie's system
by way of this criterion i.e. for independent parties $C$, $D$ and
$E$, if $S_{C|D}^{(3)}<1$, then we know that $S_{C|E}^{(3)}\geq1$.
This excludes the possibility of more than $1$ clone produced by
the teleportation process, since a second clone at $E$ would be able
to establish the same value of $S_{C|E}^{(2)}<1$ which contradicts
the monogamy result (\ref{eq:steermon3}). If EPR steering is verified
by $S_{C|B}^{(3)}<1$, then quantum teleportation of Bob's state is
verified, since any classical {}``measure and regenerate'' strategy
to generate that state would enable an infinite number of identical
states to be produced on teleportation. 

The EPR steering inequality (\ref{eq:telecond}) is thus a \emph{sufficient
condition} to demonstrate quantum teleportation. We note that the
ideal transmission of every qubit will lead to $S_{C|B}^{(3)}\rightarrow0$,
so that the quantity defined by taking the maximum of $0$ or $1-S_{C|B}^{(3)}$
gives a type of {}``figure of merit'' for the teleportation process.
This is not a true figure of merit for quantum teleportation itself,
however, as the inequality is a sufficient but not necessary condition
for quantum teleportation (as we will see in Section V). Other EPR
steering inequalities have been derived, for example, based on entropic
uncertainty relations \cite{walborn }, which could give a more effective
test of the steering.

The important point is that the three-observable steering inequality
$S_{C|B}^{(3)}<1$ is achievable at quite low inefficiencies, for
the qubit Bell state (Eq. (\ref{eq:epr1-1}) shared between Charlie
and Bob. Let $\eta_{C}$ be Charlie's efficiency and $\eta_{B}$ be
Bob's efficiency. The predictions given in Section II.C are: $(\Delta_{inf}J_{C|B}^{\theta})^{2}=\eta_{C}\{1-\eta_{B}\eta_{C}\}$
($\theta=X,Y,Z$). The Bohm EPR condition is satisfied when $\eta_{B}>1/3$
(provided $\eta_{C}>0$). The loophole-free verification is insensitive
to the losses $\eta_{C}$ of Charlie's detectors. The efficiency $\eta_{B}>1/3$
is difficult to achieve with current technology because it represents
the entire efficiency of the teleportation process, from Charlie to
Bob and including Bob\textquoteright{}s detection inefficiency. This
is because of the significant losses that take place at Alice's sending
station.

However, we can define and consider the quantum state $\rho_{CB|A}$
of Charlie and Bob, \emph{conditional} on Alice making a successful
Bell measurement and sending the classical information. In this scenario,
it is envisaged that Charlie has not made his measurement, but the
information about Alice's qubit is stored in Charlie's spin $1/2$
system. This is the case of entanglement swapping teleportation. Since
$\rho_{BC|A}$ is a quantum state, the monogamy relation is predicted
to hold for Bob and Charlie's measurements (on this conditional state),
and therefore the inequality will remain a signature of quantum teleportation.
In that case, we can argue that we have gained confirmation of quantum
teleportation regardless of inefficiencies at Alice's sending station.
The requirement of $\eta_{B}>1/3$ is determined by the efficiency
of Bob\textquoteright{}s detection and the losses on Bob\textquoteright{}s
EPR channel only. This level of efficiency has been realised in the
loophole-free steering experiments of Wittman et al \cite{wittman}
and would appear to be quite feasible. 

In many situations, the EPR channels of Alice and Bob are propagated
from a common source. To achieve true quantum teleportation for that
case, since the sensitivity is with respect to Bob's efficiency $\eta_{B}$,
the EPR source is best placed to close to Bob's station, in order
to minimise losses on Bob's channel. It also becomes essential that
he has the best detectors. These sorts of issues are discussed in
Ref. \cite{brancd-1}, from the perspective of quantum key distribution
(QKD).

\section{Secure qubit teleportation using two-setting EPR steering inequalities}

Let us consider the experiment where Bob and Charlie are able to demonstrate
that the EPR steering inequality (19) is satisfied. Then they can
confirm the quantum benchmark for teleportation. The monogamy relation
(\ref{eq:steermon3}) gives us a \emph{stronger} result: there can
be \emph{no more} than one party $E$ (other than Bob) also able to
satisfy the EPR inequality, $S_{C|E}^{(3)}<1$. On examining the proof
of the monogamy relation, we see that this follows because the EPR
inequality involves three observables, and hence there are three measurement
settings at each location. The result directly implies a level of
security of the qubit values shared by Charlie and Bob, because the
EPR inequality can only be satisfied if the variances in the inferences
of Charlie's qubit values are small enough. The inference variances
for the other parties must be large, and are quantifiable using the
monogamy relation (\ref{eq:steermon3}). 

We note we can improve the level of security, if Bob and Charlie use
\emph{two-setting} EPR steering inequalities. In that case, there
can be \emph{no} party (other than Bob) that can demonstrate the EPR
steering inequality. Two-setting inequalities for Pauli spins have
been derived in Refs. \cite{hw-np-steering-1,EPRsteering-1,pra}.
Here, we consider a two-setting EPR inequality expressed in terms
of the conditional variances. We find that EPR steering of Charlie's
system $C$ by Bob's measurements at $B$ is observed if 
\begin{equation}
(\Delta_{inf}\sigma_{C|B}^{X})^{2}+(\Delta_{inf}\sigma_{C|B}^{Y})^{2}<1\label{eq:twosettepr}
\end{equation}
The proof is presented in Ref. {[}37{]} and outlined in the Appendix.
This inequality is also a condition for Bohm's EPR paradox. Introducing
the steering parameter $S_{C|B}^{(2)}=(\Delta_{inf}\sigma_{C|B}^{X})^{2}+(\Delta_{inf}\sigma_{C|B}^{Y})^{2}$,
we can write the monogamy relation 
\begin{equation}
S_{C|B}^{(2)}+S_{C|E}^{(2)}\geq2\label{eq:steermon3-1}
\end{equation}
that follows on extending the results and definitions of (\ref{eq:steermon3}).
The relation has been derived in Ref. \cite{mon2}, and will \emph{always}
hold. Thus, if Bob can demonstrate $S_{C|B}^{(2)}<1$, then this ensures
that for any other party $E$ (Eve), it is the case that $S_{C|E}^{(2)}\geq1$,
which implies a minimum noise levels on Eve's inference of Charlie's
qubit values. Thus, the two-setting inequality $S_{C|B}^{(2)}<1$
confirms what we will call {}``secure teleportation''.

The two-setting EPR inequality (\ref{eq:twosettepr}) is derived based
on the uncertainty relation $(\Delta\sigma_{C}^{X})^{2}+(\Delta\sigma_{C}^{Y})^{2}\geq1$
for Pauli spins, which holds for any quantum state. The inequality
therefore also holds for the Schwinger spins, defined in (\ref{eq:schwinger})
but provided the outcomes for Charlie's spins (at $C$) are $\pm1$.
Assuming perfect detectors at the Charlie's station, and assuming
the system is prepared in a maximally correlated Bell state, it is
easy to show from the results of Section II.C and the Supplemental
Material {[}49{]} that the inequality (\ref{eq:twosettepr}) can be
satisfied, for any $\eta_{B}>1/2$. 

In short, the inequality $S_{C|B}^{(2)}<1$ can be used to confirm
secure teleportation, provided one assumes\emph{ trusted} detectors
at the Charlie's location, so that the fair sampling assumption is
justified at this location. In that case, security of the transmitted
state can be confirmed for up to 50\% losses in the teleportation
process (i.e. for Bob's channel and detectors).

\section{Demonstration of quantum teleportation at arbitrary efficiencies}

A set of EPR inequalities has been derived, that involve $m$ settings
\cite{hw-np-steering-1,loopholefreesteering}. These inequalities
can be expressed in a form similar to the steering inequalities (\ref{eq:bohmeff})
and (\ref{eq:twosettepr}), which we write as $S_{C|B}^{(3)}<1$ and
$S_{C|B}^{(2)}<1$. If $S_{C|B}^{(m)}<1$, then it is confirmed that
Bob can steer Charlie's system, using an $m$-setting inequality.
The exact form of $S_{C|B}^{(m)}$ is given by results in Ref. \cite{hw-np-steering-1,loopholefreesteering}.

It has been shown that the $m$-setting inequalities also satisfy
a monogamy relation \cite{mon2,monog}. If Bob and Charlie can demonstrate
an $m$-setting steering inequality, to confirm that Bob can steer
Charlie's system, then there can be no more than $m-2$ parties (other
than Bob) that can also demonstrate the $m$-setting inequality. The
realisation of the inequality $S_{C|B}^{(m)}<1$ therefore confirms
quantum teleportation. This is because a classical protocol would
enable generation of an infinite number of identical teleported states,
which in turn enables an infinite number of parties to demonstrate
the inequality, in contradiction with the monogamy result. The value
$m-2$, that gives the maximum number of parties (Eve) that can possess
a non-degraded copy of Bob's state, is an indicator of the quality
of the teleportation.

We can evaluate for what efficiencies the $m$-setting inequalities
can be satisfied, assuming Bob and Charlie share a Bell state. With
the assumption that Charlie has {}``trusted detectors'' (i.e. maximum
efficiency $\eta_{C}=1$) , it has been shown in Ref. \cite{loopholefreesteering}
that the inequality $S_{C|B}^{(m)}<1$ can be satisfied for optimal
measurement choices, provided $\eta_{B}>1/m$. This is an important
results that indicates quantum teleportation can be demonstrated \emph{for
arbitrary losses} at Bob's receiving station, provided we can make
the assumption of fair sampling at the generation (Charlie) stage.

\section{Braunstein-Kimble criticism}

Braunstein and Kimble have commented that the quality of qubit quantum
teleportation is limited by the {}``no detection'' outcomes at Bob's
location \cite{braunkimblecrit}. They also point out that the low
efficiency for generation of the EPR pair when using parametric amplification
would lead to a high incidence of {}``no detection'' outcomes, even
with ideal detectors. As mentioned by them, these problems can be
overcome e.g. by heralding the EPR pair \cite{en swp,longdistphoton}.
In terms of establishing a deterministic teleportation, high efficiency
of the teleportation process is essential {[}6{]}. 

However, it remains interesting to ask whether the claim of quantum
teleportation is compromised, in the presence of the zero detections.
We have established that quantum teleportation can in principle be
demonstrated for quite low efficiencies. Lastly, we address the second
question, for the scenario conidered in this paper. We examine the
effect of the vacuum state that arises in the parametric process that
generates the photonic EPR pair. We consider that the actual EPR resource
is the quantum state of the four-mode parametric amplifier: 
\begin{eqnarray}
|\psi\rangle & = & c_{0}|0\rangle_{a+}|0\rangle_{a-}|0\rangle_{b+}|0\rangle_{b-}\nonumber \\
 &  & +c_{1}\frac{1}{\sqrt{2}}\{|1\rangle_{a+}|0\rangle_{a-}|1\rangle_{b+}|0\rangle_{b-}\nonumber \\
 &  & +|0\rangle_{a+}|1\rangle_{a-}|0\rangle_{b+}|1\rangle_{b-}\}\label{eq:paramp}
\end{eqnarray}
Here, the contributions of terms involving modes with two or more
photons have been ignored, and therefore $|c_{0}|^{2}+|c_{1}|^{2}=1$.
For simplicity, we consider the case where there are no losses. Since
a register for Alice's Bell measurement requires a coincidence at
her detectors (for the detection of the $|\Psi_{-}\rangle$ Bell state),
we note that Alice's classical signal for teleportation go-ahead will
always be correlated with a detection of a photon at Bob's detector
in that case. This means that the vacuum state has no effect on the
calculations presented for that particular scenario.

\section{Conclusion}

The objective of this paper is to propose alternative ways to signify
the quantum teleportation of a qubit, that can be applied without
postselection. Two sorts of inequalities have been presented. The
first is a single inequality that allows confirmation of quantum teleportation
without fair sampling assumptions at either stations: where the qubit
is generated (Charlie), or where it is detected after teleportation
(Bob). This inequality involves three measurement settings and can
give a demonstration of Bohm's EPR paradox, and quantum teleportation,
for efficiencies $\eta_{C}>0$ at Charlie's station and $\eta_{B}>1/3$
at Bob's station. In this case, the quantum state that is considered
as being teleported is that conditioned on Alice's successful performance
of the Bell measurement, and the teleportation protocol is one of
entanglement swapping. The proof of quantum teleportation is based
on a proof of monogamy of the EPR paradox.

The second sort of inequality can be used where a fair sampling assumption
is made at Charlie's station, so that the outcomes of his measurements
are confined to the qubit Hilbert space. Such an assumption has been
called that of {}``trusted detectors''. In that scenario, steering
of Charlie's system by Bob (and hence quantum teleportation) can potentially
be demonstrated for arbitrary efficiency at Bob's detectors. This
conclusion is based on the results of Bennet, Evans et al \cite{loopholefreesteering},
which report steering inequalities to be violated for efficiencies
$\eta_{B}>1/m$ where $m$ is the number of measurement settings.

An important example of this second sort of inequality is a Bohm's
EPR paradox inequality for two-settings ($m=2$). This inequality
requires efficiencies of $\eta_{B}>1/2$ (for correlations based on
the maximally entangled Bell state). The useful feature of the two-setting
inequality is that a high level of security of the teleported qubit
state can be deduced. There can be no other independent party (Eve)
also able to demonstrate the inequality (apart from Bob), which implies
a minimum noise level on Eve's inferences of Charlie's qubit value.

We have seen that the EPR and steering paradoxes are useful to demonstrate
quantum teleportation in the presence of loss. Are there other advantages?
A possible response relates to the nature of the derivation of the
EPR steering inequalities. It is assumed that Charlie's measurements
are of a quantum spin system, and are therefore constrained by quantum
mechanics. However, the EPR steering inequalities are derived with
\emph{no} similar assumption about what or how measurements are made
by Bob (or Eve) \cite{belldeindsec-1}. In this way, we see that the
conditions for quantum teleportation (and for the security of the
teleported state) have the advantages of being {}``one-sided device-independent''
\cite{brancd-1}.
\begin{acknowledgments}
This research was supported by an Australian Research Council Discovery
grant. I thank B Wittmann, A Zeilinger, Q. He and P Drummond for discussions
relating to steering that helped motivate this work and Run Yan Teh
for discussions on entanglement swapping.
\end{acknowledgments}

\section*{Appendix}

We give a derivation of the EPR paradox and steering inequalities.
This type of proof has been given in Ref. {[}37{]}. We begin with
the definition of the Local Hidden State (LHS) model \cite{hw-steering-1,hw2-steering-1}.
To prove steering, we need to falsify a description of the statistics
based on a LHS model, where the averages are given as
\begin{eqnarray}
\langle X_{B}X_{A}\rangle & = & \int_{R}P_{R}\langle X_{B}\rangle_{R}\langle X_{A}\rangle_{R,\rho}\label{eq:lhs}
\end{eqnarray}
Here $\int_{R}P_{R}=1$ and the $\rho$ subscript denotes that the
averages are consistent with those of a quantum density matrix. No
such constraint is made for the moments $\langle X_{i}\rangle_{R}$,
written without the subscript $\rho$. This model is one in which
the system is a probabilistic mixture of states symbolised by $R$,
with probabilities $P_{R}$. The states symbolised by $R$ (without
the subscript $\rho$) may be identified as the local hidden variable
states assumed in Bell's Local Hidden Variable models. The summation
over all possible states $R$ can be denoted either by an integral
or by a discrete summation, similar to the situation for Bell's Local
Hidden Variable models \cite{Bell}. 

The average conditional uncertainty is
\begin{equation}
(\Delta_{inf}\sigma_{A}^{X})^{2}=\sum_{x_{j}^{B}}P(x_{j}^{B})\{\Delta(\sigma_{A}^{X}|x_{j}^{B})\}^{2}\label{eq:inf-2}
\end{equation}
where we denote the possible results of the specified measurement
at $B$ by $\{x_{j}^{B}\}$. Using the definitions, and assuming the
mixtures as implied by the LHS model, we see step by step that:
\begin{eqnarray*}
\sum_{x_{j}^{B}}P(x_{j}^{B})\{\Delta(\sigma_{A}^{X}|x_{j}^{B})\}^{2}\,\,\,\,\,\,\,\,\,\,\,\,\,\,\,\,\,\,\,\,\,\,\,\,\,\,\,\,\,\,\,\,\,\,\,\,\,\,\,\,\,\,\,\,\,\,\,\,\,\,\,\,\,\,\,\,\,\,\,\,\,\,\,\,\,\,\,\,\\
=\sum_{x_{j}^{B}}P(x_{j}^{B})\sum_{\sigma_{x}^{A}}P(\sigma_{A}^{X}|x_{j}^{B})\{\sigma_{A}^{X}-\langle\sigma_{A}^{X}|x_{j}^{B}\rangle\}^{2}\,\,\,\,\,\,\,\,\,\,\,\,\,\,\,\,\,\,\,\,\,\,\,\,\,\,\,\,\,\,\,\,\,\,\,\,\,\,\,\,\\
=\sum_{x_{j}^{B},\sigma_{A}^{X}}P(x_{j}^{B},\sigma_{A}^{X})\{\sigma_{A}^{X}-\langle\sigma_{A}^{X}|x_{j}^{B}\rangle\}^{2}\,\,\,\,\,\,\,\,\,\,\,\,\,\,\,\,\,\,\,\,\,\,\,\,\,\,\,\,\,\,\,\,\,\,\,\,\,\,\,\,\,\,\,\,\,\,\,\,\,\,\,\\
=\sum_{R}P_{R}\sum_{x_{j}^{B},\sigma_{A}^{X}}P_{R}(x_{j}^{B},\sigma_{A}^{X})\{\sigma_{A}^{X}-\langle\sigma_{A}^{X}|x_{j}^{B}\rangle\}^{2}\,\,\,\,\,\,\,\,\,\,\,\,\,\,\,\,\,\,\,\,\,\,\,\,\,\,\,\,\,\,\,\,\,\\
\geq\sum_{R}P_{R}\sum_{x_{j}^{B}}P_{R}(x_{j}^{B})\sum_{\sigma_{A}^{X}}P_{R}(\sigma_{A}^{X}|x_{j}^{B})\,\,\,\,\,\,\,\,\,\,\,\,\,\,\,\,\,\,\,\,\,\,\,\,\,\,\,\,\,\,\,\,\,\,\,\,\,\,\,\,\,\,\,\,\,\,\,\,\,\,\,\,\,\,\,\,\,\,\,\,\\
\times\{\sigma_{A}^{X}-\langle\sigma_{A}^{X}|x_{j}^{B}\rangle_{R}\}^{2}\,\,\,\,\,\,\,\,\,\,\,\,\,\,\,\,\,\,\,\,\,\,\,\,\,\,\,\,\,\,\\
=\sum_{R}P_{R}\sum_{x_{j}^{B}}P_{R}(x_{j}^{B})\{\Delta_{R}(\sigma_{A}^{X}|x_{j}^{B})\}^{2}\,\,\,\,\,\,\,\,\,\,\,\,\,\,\,\,\,\,\,\,\,\,\,\,\,\,\,\,\,\,\,\,\,\,\,\,\,\,\,\,\,\,\,\,\,\,\,\,\,\,\,\,\,\,\,\,\,\,\,\,\,\,\,\,\,\,\,\,\,\,\,\\
=\sum_{R}P_{R}\{\Delta_{inf,R}\sigma_{A}^{X}\}^{2}\,\,\,\,\,\,\,\,\,\,\,\,\,\,\,\,\,\,\,\,\,\,\,\,\,\,\,\,\,\,\,\,\,\,\,\,\,\,\,\,\,\,\,\,\,\,\,\,\,\,\,\,\,\,\,\,\,\,\,\,\,\,\,\,\,\,\,\,\,\,\,\,\,\,\,\,\,\,\,\,\,\,\,\,\,\,\,\,\,\,\,\,\,\,\,\,\,\,\,\,\,\,\,\,\,
\end{eqnarray*}
The fourth line follows using that for a probabilistic mixture $P(x_{j}^{B},\sigma_{X}^{A})=\sum_{R}P_{R}P_{R}(x_{j}^{B},\sigma_{X}^{A})$.
The fifth line follows from the fact that $\langle(x-\delta)^{2}\rangle\geq\langle(x-\langle x\rangle)^{2}\rangle$
where $\delta$ is any number. Here, the subscripts $R$ imply that
the probabilities, averages and variances are with respect to the
state $R$. Now, if we assume the separability between the bipartition
$A-B$ for each state $R$, in accordance with the LHS model, then
\begin{equation}
P_{R}(x_{j}^{B},\sigma_{A}^{X})=P_{R}(x_{j}^{B})P_{R}(\sigma_{A}^{X})\label{eq:sep}
\end{equation}
This implies $\langle\sigma_{A}^{X}|x_{j}^{B}\rangle_{R}=\langle\sigma_{A}^{X}\rangle$
and $\{\Delta_{R}(\sigma_{A}^{X}|x_{j}^{B})\}^{2}=\{\Delta_{R}(\sigma_{A}^{X})\}^{2}$.
Then we find, on using $\sum_{x_{j}^{B}}P_{R}(x_{j}^{B})=1$, that
we can write $ $$\{\Delta_{inf,R}\sigma_{A}^{X}\}^{2}=\{\Delta_{R}(\sigma_{A}^{X})\}^{2}$.
Thus, 
\begin{eqnarray*}
(\Delta_{inf}\sigma_{A}^{X})^{2}+(\Delta_{inf}\sigma_{A}^{Y})^{2}\geq\,\,\,\,\,\,\,\,\,\,\,\,\,\,\,\,\,\,\,\,\,\,\,\,\,\,\,\,\\
\sum_{R}P_{R}(\{\Delta_{R}(\sigma_{A}^{X})\}^{2}+\{\Delta_{R}(\sigma_{A}^{Y})\}^{2})
\end{eqnarray*}
and
\begin{eqnarray*}
(\Delta_{inf}\sigma_{A}^{X})^{2}+(\Delta_{inf}\sigma_{A}^{Y})^{2}+(\Delta_{inf}\sigma_{A}^{Z})^{2}\geq\,\,\,\,\,\,\,\,\,\,\,\,\,\,\,\,\,\,\,\,\\
\sum_{R}P_{R}(\{\Delta_{R}(\sigma_{A}^{X})\}^{2}+\{\Delta_{R}(\sigma_{A}^{Y})\}^{2}+\{\Delta_{R}(\sigma_{A}^{Z})\}^{2})
\end{eqnarray*}
Because in the LHS model (\ref{eq:lhs}) we assume the states at $A$
are local quantum states, we can use quantum uncertainty relations
to derive a final steering inequality: e.g. $\{\Delta_{R}(\sigma_{A}^{X})\}^{2}+\{\Delta_{R}(\sigma_{A}^{Y})\}^{2}\geq1$
for any quantum state, and hence the LHS model implies
\begin{eqnarray}
(\Delta_{inf}\sigma_{A}^{X})^{2}+(\Delta_{inf}\sigma_{A}^{Y})^{2} & \geq & 1\label{eq:3}
\end{eqnarray}
Also, the uncertainty relation (\ref{eq:spinun-1}) will hold for
any quantum state. Thus the LHS model implies 
\begin{eqnarray}
(\Delta_{inf}S_{A}^{X})^{2}+(\Delta_{inf}S_{A}^{Y})^{2}+(\Delta_{inf}S_{A}^{Z})^{2}\nonumber \\
\geq\langle n_{A}^{2}\rangle-\langle n_{A}\rangle^{2}+2\langle n_{A}\rangle\label{eq:bohmeff-1}
\end{eqnarray}
Violation of either of these two inequalities implies failure of the
LHS model, and therefore implies steering of $A$ by $B$. The violation
will also imply an EPR paradox in each case, because the inferred
uncertainties represent the uncertainties of the {}``elements of
reality'', that exist to describe the local state of $A$, according
to the EPR premises of local realism. These uncertainties are not
compatible with the quantum uncertainty relation.

\end{document}